# CLUSTER BASED COST EFFICIENT INTRUSION DETECTION SYSTEM FOR MANET

Saravanan Kumarasamy[1], Hemalatha B[2] and Hashini P[3]
[1] Asst. Professor/CSE Erode Sengunthar Engineering College, Erode, India
[2, 3] B.E. C.S.E. Student, Erode Sengunthar Engineering College, Erode, India
Email:saravanankumarasamy@gmail.com

**ABSTRACT**

Mobile ad-hoc networks are temporary wireless networks. Network resources are abnormally consumed by intruders. Anomaly and signature based techniques are used for intrusion detection. Classification techniques are used in anomaly based techniques. Intrusion detection techniques are used for the network attack detection process. Two types of intrusion detection systems are available. They are anomaly detection and signature based detection model. The anomaly detection model uses the historical transactions with attack labels. The signature database is used in the signature based IDS schemes.

The mobile ad-hoc networks are infrastructure less environment. The intrusion detection applications are placed in a set of nodes under the mobile ad-hoc network environment. The nodes are grouped into clusters. The leader nodes are assigned for the clusters. The leader node is assigned for the intrusion detection process. Leader nodes are used to initiate the intrusion detection process. Resource sharing and lifetime management factors are considered in the leader election process. The system optimizes the leader election and intrusion detection process.

The system is designed to handle leader election and intrusion detection process. The clustering scheme is optimized with coverage and traffic level. Cost and resource utilization is controlled under the clusters. Node mobility is managed by the system.

## 1. INTRODUCTION

Unlike traditional networks, the Mobile Ad hoc Networks (MANETs) have no fixed chokepoints/bottlenecks where Intrusion Detection Systems (IDSs) can be deployed [3]. Hence, a node may need to run its own IDS [1] and cooperate with others to ensure security. This is very inefficient in terms of resource consumption since mobile nodes are energy-limited. To overcome this problem, a common approach is to divide the MANET into a set of 1-hop clusters where each node belongs to at least one cluster. The nodes in each cluster elect a leader node (cluster head) to serve as the IDS for the entire cluster.

The leader-IDS election process can be either random or based on the connectivity. Both approaches aim to reduce the overall resource consumption of IDSs in the network. However, we notice that nodes usually have different remaining resources at any given time, which should be taken into account by an election scheme. Unfortunately, with the random model, each node is equally likely to be elected regardless of its remaining resources [11]. The connectivity index-based approach elects a node with a high degree of connectivity even though the node may have little resources left. With both election schemes, some nodes will die faster than others, leading to a loss in connectivity and potentially the partition of network. Although it is clearly desirable to balance the resource consumption of IDSs among nodes, this objective is difficult to achieve since the resource level is the private information of a node. Unless sufficient incentives are provided, nodes might misbehave by acting selfishly and lying about their resources level to not consume their resources for serving others while receiving others services. Moreover, even when all nodes can truthfully reveal their resource levels, it remains a challenging issue to elect an optimal collection of leaders to balance the overall resource consumption without flooding the network. Next, we motivate further discussions through a concrete example.

The problem of selfishness and energy balancing exists in many other applications to which our solution is also applicable. Like in IDS scheme, leader election is needed for routing and key distribution [6] in MANET. In key management, a central key distributor is needed to update the keys of nodes. In routing, the nodes are grouped into small clusters and each cluster elects a cluster head (leader) to forward the packets of other nodes. Thus, one node can stay alive, while others can be in the energy-saving mode. The election of a leader node is done randomly, based on connectivity (nodes' degree) or based on a node's weight. We have already pointed out the problems of random model and connectivity model. We believe that a weightbased leader election should be the proper method for election. Unfortunately, the information regarding the remaining energy is private to a node, and thus, not verifiable. Since nodes might behave selfishly, they might lie about their resource level to avoid being the leader if there is no mechanism to motivate them. Our method can effectively address this issue.

## 2. RELATED WORK

This section reviews related work on intrusion detection in MANET, the application of mechanism design to networks, and the application of leader election scheme to routing and key distribution.

### 2.1. INTRUSION DETECTION SYSTEMS IN MANET

The difference between wired infrastructure networks and mobile ad hoc networks raises the need for new IDS models that can handle new security challenges. Due to the security needs in MANET, a cooperative intrusion detection model has been proposed, where every node participates in running its IDS in order to collect and identify possible intrusions. If an anomaly is detected with weak evidence, then a global detection process is initiated for further investigation about the intrusion through a secure channel. An extension of this model was proposed, where a set of intrusions can be identified with their corresponding sources. Moreover, the authors address the problem of runtime resource constraints through modeling a repeatable and random leader election framework. An elected leader is responsible for detecting intrusions for a predefined period of time. Unlike our work, the random election scheme does not consider the remaining resources of nodes or the presence of selfish nodes.

A modular IDS system based on mobile agents is proposed and the authors point out the impact of limited computational and battery power on the network monitoring tasks. Again, the solution ignores both the difference in remaining resources and the selfishness issue. To motivate the selfish nodes in routing, CONFIDANT proposes a reputation system where each node keeps track of the misbehaving nodes. The reputation system is built on the negative evaluations rather than positive impression. Whenever a specific threshold is exceeded, an appropriate action is taken against the node. Therefore, nodes are motivated to participate by punishing the misbehaving ones through giving a negative reputation. As a consequence of such a design, a malicious node can broadcast a negative impression about a node in order to be punished. On the other hand, CORE [2] is proposed as a cooperative enforcement mechanism based on monitoring and reputation systems. The goal of this model is to detect selfish nodes and enforce them to cooperate. Each node keeps track of other nodes cooperation using reputation as a metric. CORE ensures that misbehaving nodes are punished by gradually excluding them from communication services. In this model, the reputation is calculated based on data monitored by local nodes and information provided by other nodes involved in each operation. In contrast to such passive approaches, our solution proactively encourages nodes to behave honestly through computing reputations based on mechanism design. Moreover, it is able to punish misbehaving leaders through a cooperative punishment system based on cooperative game theory. In addition to this, a noncooperative game is designed to help the leader IDS to increase the probability of detection by distributing the node's sampling over the most critical links.

### 2.2. APPLICATION OF MECHANISM DESIGN

As a subfield of microeconomics and game theory, mechanism design has received extensive studies in microeconomics for modeling economical activities. Nisan and Ronen apply mechanism design for solving the least-cost path and task scheduling problem. Distributed mechanism design based on VCG is first introduced in a direct extension of Border Gateway Protocol (BGP) for computing the lowest cost routes. Moreover, the authors outlined the basics of distributed mechanism design and reviewed the results done on multicast cost sharing and interdomain routing. Mechanism design has been used for routing purposes in MANETs, such as a truthful ad-hoc-VCG mechanism for finding the most cost-efficient route in the presence of selfish nodes. In [9], the authors provide an incentive compatible auction scheme to enable packet forwarding services in MANETs using VCG; a continuous auction process is used to determine the distribution of bandwidth, and incentives are given as monetary rewards. To our best knowledge, this work is among the first efforts in applying mechanism design theory to address the security issues in MANETs, in particular, the leader election for intrusion detection. This paper is the extension of [4], where we presented the leader election mechanism in a static environment without addressing different performance overhead.

## 2.3. LEADER ELECTION APPLICATIONS

Distributed algorithms for clustering and leader election have been addressed in different research work [5]. These algorithms can be classified into two categories: Cluster-first or leader-first. In the cluster first approach, a cluster is formed, and then, the nodes belonging to that cluster elect a leader node. In the leader first approach, a set of leader nodes is elected first, then the other nodes are assigned to different leader nodes. Some of the methods assume that there exist a weight associated with each node or there exist a trusted authority to certify each node's metric (weight) which is used to elect a leader. We consider these assumptions as quite strong for MANET. Our model is able to run in a clustered and nonclustered network where we are able to perform better results with respect to different performance metrics.

## 3. PROBLEM STATEMENT

We consider an MANET where each node has an IDS and a unique identity. To achieve the goal of electing the most cost-efficient nodes as leaders in the presence of selfish and malicious nodes, the following challenges arise: First, the resource level that reflects the cost of analysis is considered as private information. As a result, the nodes can reveal fake information about their resources if that could increase their own benefits. Second, the nodes might behave normally during the election but then deviate from normal behavior by not offering the IDS service to their voted nodes.

In our model, we consider MANET as an undirected graph $G = (N,L)$, where $N$ is the set of nodes and $L$ is the set of bidirectional links. We denote the cost of analysis vector as $C = \{c_1, c_2,...c_n\}$ where $n$ is the number of nodes in $N$. We denote the election process as a function $vt_k(C, i)$, where $vt_k(C, i) = 1$ if a node $i$ votes for a node $k$; $vt_k(C, i) = 0$, otherwise. We assume that each elected leader allocates the same budget $B$ for each node that has voted for it. Knowing that the total budget will be distributed among all the voting nodes according to their reputation. This will motivate the nodes to cooperate in every election round that will be held on every time $T_{ELECT}$. Thus, the model will be repeatable. For example, if $B = 25$ packet/sec and the leader gets three votes, then the leader's sampling budget is 75 packet/sec. This value is divided among the three nodes based on their reputation value. The objective of minimizing the global cost of analysis while serving all the nodes can be expressed by the following Social Choice Function (SCF):

$$\text{SCF} = S(C) = \min \sum_{k \in N} c_k \cdot \left( \sum_{i \in N} vt_k(C,i) \cdot B \right)$$

Clearly, in order to minimize this SCF, the following must be achieved. First, we need to design incentives for encouraging each node in revealing its true cost of analysis value $c$. Second, we need to design an election algorithm that can provably minimize the above SCF while not incurring too much of the performance overhead.

## 4. LEADER ELECTION MECHANISM

### 4.1 MECHANISM DESIGN BACKGROUND

Mechanism design is a subfield of microeconomics and game theory. Mechanism design uses game theory tools to achieve the desired goals. The main difference between game theory and mechanism design is that the former can be used to study what could happen when independent players act selfishly. On the other hand, mechanism design allows a game designer to define rules in terms of the SCF such that players will play according to these rules. The balance of IDS resource consumption problem can be modeled using mechanism design theory with an objective function that depends on the private information of the players. In our case, the private information of the player is the cost of analysis which depends on the player's energy level. Here, the rational players select to deliver the untruthful or incomplete information about their preferences if that leads to individually better outcomes [10]. The main goal of using mechanism design [7] is to address this problem by: 1) designing incentives for players (nodes) to provide truthful information about their preferences over different Outcomes and 2) computing the optimal system-wide solution, which is defined according to (1).

A mechanism design model consists of n agents where each agent $i \in \{1, \ldots, n\}$ has a private information, $\phi_i i \in \theta_i$, known as the agent's type. Moreover, it defines a set of strategies $A_i$ for each agent i. The agent can choose any strategy $a_i \in A_i$ to input in the mechanism. According to the inputs $(a_i, \ldots, a_n)$ of all the agents, the mechanism calculates an output $o = o(a_1, \ldots, a_n)$ and payment vector $p = (p_1, \ldots, p_n)$, where $p_i = p_i(a_1, \ldots, a_n)$. The preference of each agent from the output is calculated by a valuation function $v_i(\theta_i, o)$. This is a quantification in terms of a real number to evaluate the output for an agent i. Thus, the utility of a node is calculated as $u_i = p_i - v_i(\theta_i, o)$. This means that the utility is the combination of output measured by valuation function and the payment it receives from the mechanism.

In direct revelation mechanism, every agent i has a type $\theta_i$. Each agent gives an input $a_i(\theta_i)$ to the mechanism. The agent chooses the strategy according to its type, where $a_i(\theta_i) = \_i$, which is chosen from the strategy set $\theta =$ {Selfish; Normal}. We assume that normal agents follow the protocol, whereas selfish agents deviate from the defined protocol if the deviation leads to a higher utility. Although the prime objective of these agents is not to actively harm others but their presence can passively harm others.

Last but not least, the mechanism provides a global output from the input vector and also computes a specific payment for each agent. The goal is to design a strategy proof mechanism where each agent gives an input based on its real type $\theta_i$ such that it maximizes its utility regardless of the strategies of others. A strategy is dominated by another strategy if the second strategy is at least as good as the other one regardless of the other players' strategy. This is expressed as follows:

$$p_i - v_i(\theta_i^*, o) = u_i^* \geq u_i = p_i - v_i(\theta_i, o),$$

where $\theta_i^*$ denotes nonselfishness and $\theta_i$ denotes selfishness. Note that ui is maximized only when pi is given by the mechanism. The question is: How to design the payments in a way that makes truth-telling the dominant strategy? In other words, how to motivate nodes to reveal truthfully their valuation function $v_i(\theta_i^*, o)$? The VCG mechanism answers this question by giving the nodes a fixed payment independent of the nodes' valuation, which is equal to the second best valuation. The design of the payment, according to our scenarios, is given in the following sections. A general overview of mechanism design can be found in [8].

### 4.2. THE MECHANISM MODEL

We treat the IDS resource consumption problem as a game where the N mobile nodes are the agents/players. Each node plays by revealing its own private information which is based on the node's type $\theta_i$. The type $\theta_i$ is drawn from each player's available type set $\theta_i = \{Normal, Selfish\}$. Each player selects his own strategy/type according to how much the node values the outcome. If the player's strategy is normal, then the node reveals the true cost of analysis. We assume that each player i has a utility function:

$$u_i(\theta_i) = p_i - v_i(\theta_i, o(\theta_i, \theta_{-i})), \quad - \quad (2)$$

where .

- $\theta_{-i}$ is the type of all the other nodes except i.
- $v_i$ is the valuation of player i of the output $o \in O$, knowing that O is the set of possible outcomes. In our case, if the node is elected, then vi is the cost of analysis $c_i$. Otherwise, $v_i$ is 0 since the node will not be the leader, and hence, there will be no cost to run the IDS.
- $p_i \in \Re$ is the payment given by the mechanism to the elected node. Payment is

given in the form of reputation. Nodes that are not elected receive no payment.

Note that $u_i(\theta_i)$ is what the player usually seeks to maximize. It reflects the amount of benefits gained by player i if he follows a specific type $\theta_i$. Players might deviate from revealing the truthful valuation for the cost of analysis if that could lead to a better payoff. Therefore, our mechanism must be strategy-proof where truth-telling is the dominant strategy. To play the game, every node declares its corresponding cost of analysis where the cost vector C is the input of our mechanism. For each input vector, the mechanism calculates its corresponding output o = o($\theta_i$, . . . , $\theta_n$) and a payment vector p = ($p_1$, . . . ,$p_n$). Payments are used to motivate players to behave in accordance with the mechanism goals. In the following sections, we will formulate the following components:
1. Cost of analysis function: It is needed by the nodes to compute the valuation function.
2. Reputation system: It is needed to show how:
    a. Incentives are used once they are granted.
    b. Misbehaving nodes are catched and punished.
3. Payment design: It is needed to design the amount of incentives that will be given to the nodes based on VCG.

## 4.3 COST OF ANALYSIS FUNCTION

During the design of the cost of analysis function, the following two problems arise: First, the energy level is considered as private and sensitive information and should not be disclosed publicly. Such a disclosure of information can be used maliciously for attacking the node with the least resources level. Second, if the cost of analysis function is designed only in terms of nodes' energy level, then the nodes with the low energy level will not be able to contribute and increase their reputation values.

To solve the above problems, we design the cost of analysis function with the following two properties: Fairness and Privacy. The former is to allow nodes with initially less resources to contribute and serve as leaders in order to increase their reputation. On the other hand, the latter is needed to avoid the malicious use of the resources level, which is considered as the most sensitive information. To avoid such attacks and provide fairness, the cost of analysis is designed based on the reputation value, the expected number of time slots that a node wants to stay alive in a cluster, and energy level. Note that the expected number of slots and energy level are considered as the nodes' private information.

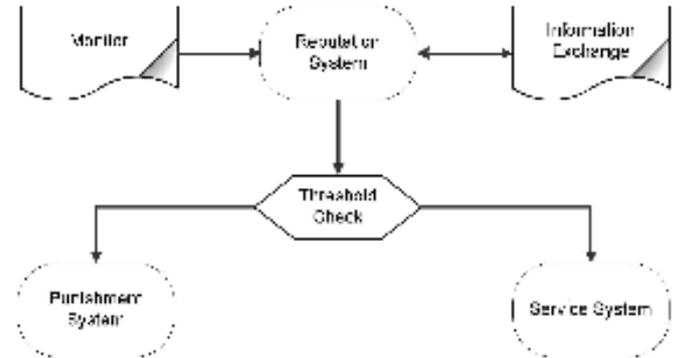

Fig. 4.3.1. Reputation system model.

## 4.4. REPUTATION SYSTEM MODEL

Before we design the payment, we need to show how the payment in the form of reputation can be used to: 1) motivate nodes to behave normally and 2) punish the misbehaving nodes. Moreover, it can be used to determine whom to trust. To motivate the nodes in behaving normally in every election round, we relate the cluster's services to nodes' reputation. This will create a competition environment that motivates the nodes to behave normally by saying the truth. To enforce our mechanism, a punishment system is needed to prevent nodes from behaving selfishly after the election. Misbehaving nodes are punished by decreasing their reputation, and consequently, are excluded from the cluster services if the reputation is less than a predefined threshold. As an extension to our model, we can extend our reputation system to include different sources of information such as routing and key distribution with different assigned weights.

## 4.5. CILE PAYMENT DESIGN

In CILE, each node must be monitored by a leader node that will analyze the packets for other ordinary nodes. Based on the cost of analysis vector C, nodes will cooperate to elect a set of leader nodes that will be able to analyze the traffic across the whole network and handle the monitoring process. This increases the efficiency and balances the resource consumption of an IDS in the network. Our mechanism provides payments to the elected leaders for serving others. The payment is based on a per-packet price that depends on the number of votes the elected nodes get. The nodes that do not get any vote from others will not receive any payment. The payment is in the form of reputations, which are then used to allocate the leader's sampling budget for each node. Hence, any node will strive to increase its reputation in order to receive more IDS services from its corresponding leader.

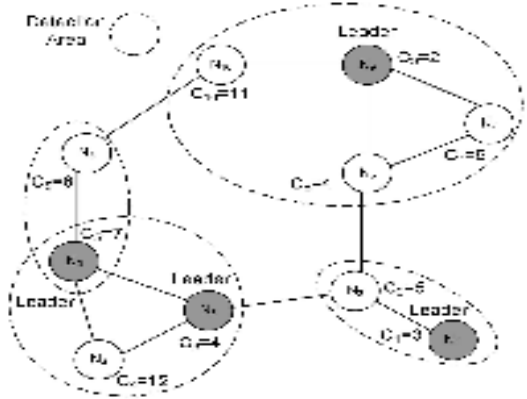

**Fig 4.5.1. An example of leader election.**

### 4.6. CDLE PAYMENT DESIGN

In CDLE, the whole network is divided into a set of clusters where a set of 1-hop neighbor nodes forms a cluster. Here, we use the scheme to cluster the nodes into 1-hop clusters. Each cluster then independently elects a leader among all the nodes to handle the monitoring process based on nodes' analysis cost. Our objective is to find the most cost-efficient set of leaders that handle the detection process for the whole network. Hence, our social choice function is still as in (1). To achieve the desired goal, payments are computed using the VCG mechanism where truth-telling is proved to be dominant. Like CILE, CDLE provides payment to the elected node and the payment is based on a per-packet price that depends on the number of votes the elected node gets.

### 5. LEADER ELECTION ALGORITHM

To run the election mechanism, we propose a leader election algorithm that helps to elect the most cost-efficient leaders with less performance overhead compared to the network flooding model. We devise all the needed messages to establish the election mechanism taking into consideration cheating and presence of malicious nodes. Moreover, we consider the addition and removal of nodes to/from the network due to mobility reasons. Finally, the performance overhead is considered during the design of the given algorithm where computation, communication, and storage overhead are derived.

### 5.1. OBJECTIVES AND ASSUMPTIONS

To design the leader election algorithm, the following requirements are needed: 1) To protect all the nodes in a network, every node should be monitored by a leader and 2) to balance the resource consumption of IDS service, the overall cost of analysis for protecting the whole network is minimized. In other words, every node has to be affiliated with the most cost-efficient leader among its neighbors. Our algorithm is executed in each node taking into consideration the following assumptions about the nodes and the network architecture:

- Every node knows its (2-hop) neighbors, which is reasonable since nodes usually maintain a table about their neighbors for routing purposes.
- Loosely synchronized clocks are available between nodes.
- Each node has a key pair for establishing a secure communication between nodes.
- Each node is aware of the presence of a new node or removal of a node.

For secure communication, we can use a combination of TESLA and public key infrastructure. With the help of TESLA, loosely synchronized clocks can be available. Nodes can use public key infrastructure during election and TESLA in other cases. Recent investigations showed that computationally limited mobile nodes can also perform public key operations.

### 5.2. LEADER ELECTION

To start a new election, the election algorithm uses four types of messages. Hello, used by every node to initiate the election process; Begin-Election, used to announce the cost of a node; Vote, sent by every node to elect a leader; and Acknowledge, sent by the leader to broadcast its payment, and also as a confirmation of its leadership. For describing the algorithm, we use the following notation:

- service-table(k): The list of all ordinary nodes, those voted for the leader node k.

- reputation-table(k): The reputation table of node k. Each node keeps the record of reputation of all other nodes.
- neighbors(k): The set of node k's neighbors.
- leadernode(k): The ID of node k's leader. If node k is running its own IDS, then the variable contains k.
- leader(k): A boolean variable that sets to TRUE if node k is a leader and FALSE otherwise.

Initially, each node k starts the election procedure by broadcasting a Hello message to all the nodes that are 1 hop from node k and starts a timer $T_1$. This message contains the hash value of the node's cost of analysis and its unique identifier (ID). This message is needed to avoid cheating where further analysis is conducted.

**Algorithm 1 (Executed by every node)**
/* On receiving Hello, all nodes reply with their cost */
1. if (received Hello from all neighbors) then
2.     Send Begin-Election ($ID_k$; $cost_k$);
3. else if(neighbors(k) = $\phi$) then
4.     Launch IDS.
5. end if

On expiration of $T_1$, each node k checks whether it has received all the hash values from its neighbors. Nodes from whom the Hello message have not received are excluded from the election. On receiving the Hello from all neighbors, each node sends Begin-Election as in Algorithm 1, which contains the cost of analysis of the node, and then, starts timer $T_2$. If node k is the only node in the network or it does not have any neighbors, then it launches its own IDS.

**Algorithm 2 (Executed by every node)**
/* Each node votes for one node among the neighbors */
1. if ($\forall$ n $\varepsilon$ neighbor(k), $\exists$ i $\varepsilon$ n : $c_i \leq c_n$) then
2.     send Vote($ID_k$, $ID_i$, $cost_{j \neq i}$);
3.     leadernode(k) := i;
5. end if

On expiration of $T_2$, the node k compares the hash value of Hello to the value received by the Begin-Election to verify the cost of analysis for all the nodes. Then, node k calculates the least-cost value among its neighbors and sends Vote for node i as in Algorithm 2. The Vote message contains the $ID_k$ of the source node, the $ID_i$ of the proposed leader, and second least cost among the neighbors of the source node $cost_{j \neq i}$. Then, node k sets node i as its leader in order to update later on its reputation. Note that the second least cost of analysis is needed by the leader node to calculate the payment. If node k has the least cost among all its neighbors, then it votes for itself and starts timer $T_3$.

**Algorithm 3 (Executed by Elected leader node)**
/* Send Acknowledge message to the neighbor nodes */
1. Leader(i) := TRUE;
2. Compute Payment, $P_i$;
3. update$_{service-table}$(i);
4. update$_{reputation-table}$(i);
5. Acknowledge = Pi + all the votes;
6. Send Acknowledge (i);
7. Launch IDS.

On expiration of $T_3$, the elected node i calculates its payment using (5) and sends an Acknowledge message to all the serving nodes as in Algorithm 3. The Acknowledge message contains the payment and all the votes the leader received. The leader then launches its IDS.

Each ordinary node verifies the payment and updates its reputation table according to the payment. All the messages are signed by the respective source nodes to avoid any kind of cheating. At the end of the election, nodes are divided into two types: Leader and ordinary nodes. Leader nodes run the IDS for inspecting packets, during an interval $T_{ELECT}$, based on the relative reputations of the ordinary nodes. We enforce reelection every period $T_{ELECT}$ since it is unfair and unsafe for one node to be a leader forever. Even if the topology remains same after $T_{ELECT}$ time, all the nodes go back to initial stage and elect a new leader according to the above algorithms.

## 6. ENERGY EFFICIENT LEADER ELECTION MODEL

The system is designed to handle leader election and intrusion detection process. The clustering scheme is optimized with coverage and traffic level. Cost and resource utilization is controlled under the clusters. Node mobility is managed by the system. The system is designed to perform intrusion detection on MANET environment with energy management. Intrusion detection operations are initiated by the leader nodes. The anomaly based model is used for the intrusion detection process. The system is divided into five major modules. They are Network analysis, Clustering process, Leader Election, Detector assignment and Intrusion Detection.

The network analysis module is designed to find out the neighbor details. Clustering process module is designed to group up the nodes. Cluster leader node is selected with resource levels. Detector assignment module is designed to place intrusion detector application. The intrusion detection process is performed with network transactions.

## 6.1. NETWORK ANALYSIS

The network node status information are collected in this module. Neighbor nodes and their resource levels are collected and updated. Computational and storage details are also collected from the nodes. Network status details are periodically updated.

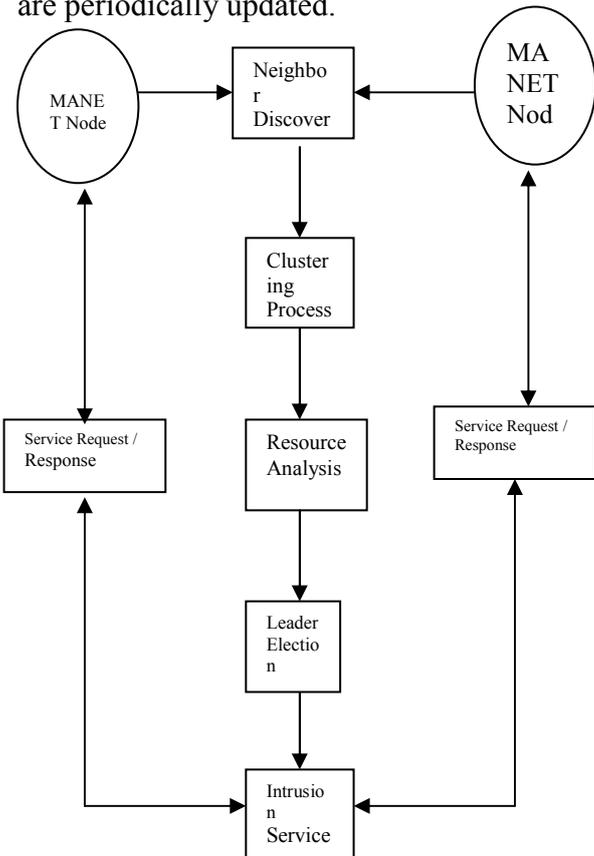

**Fig.6.2.1.Intrusion Detection System for MANET**

## 6.2. CLUSTERING PROCESS

The neighborhood nodes are grouped into clusters. The clusters are formed with coverage and traffic information. The cluster details are updated in intervals. The cluster resources are shared by the nodes.

## 6.3. LEADER ELECTION

The leader is elected for each cluster. The resource and energy details are considered in the leader selection process. The leader nodes are assigned with incentives. Reputation is provided with reference to incentives.

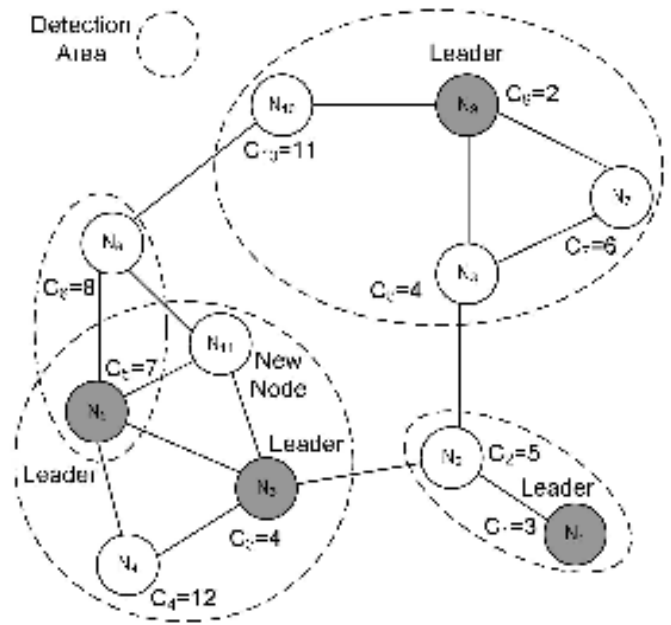

**Fig. 6.2.2. An MANET after adding a new node.**

## 6.4. DETECTOR ASSIGNMENT

The detector responds for the intrusion detection process. The detector is placed with reference to the resource levels. The leader node is assigned with detector module. The detector collects all transactions for intrusion detection process.

## 6.5. INTRUSION DETECTION

The intrusion detection process is initiated in intervals. Dynamic interval estimation is used in the system. The Bayesian classification technique is used for intrusion detection process. The intrusion detection results are passed to the requested node.

## 7. CONCLUSION

The mobile ad-hoc networks are infrastructure less environment. Leader nodes are used to initiate the intrusion detection process. Resource sharing and lifetime management factors are considered in the leader election process. The system optimizes the leader election and intrusion detection process. The system reduces the energy consumption. Network traffic is reduced by the system. Dynamic interval is assigned for intrusion detection process.